\documentclass[12pt]{amsart}
\usepackage{geometry} 
\usepackage{epsfig}
\usepackage{graphicx}

\title{Concurrence Topology of Some Cancer Genomics Data}
\author{Steven P. Ellis}

\address{Unit 42, NYSPI \\
1051 Riverside Dr. \\
New York, NY 10032 \\
U.S.A.}

\email{spe4ellis@aol.com}

\date{9/6/2017} 

\thanks{This research is supported in part by NIH grant U 54 CA193313.}


\begin{document}

\maketitle

\begin{abstract}
The topological data analysis method ``concurrence topology'' is applied to mutation frequencies in 69 genes in  glioblastoma data. In dimension 1 some apparent ``mutual exclusivity'' is found. By simulation of data having approximately the same second order dependence structure as that found in the data, it appears that one triple of mutations, PTEN, RB1, TP53, exhibits mutual exclusivity that depends on special features of the third order dependence and may reflect global dependence among a larger group of genes. A bootstrap analysis suggests that this form of mutual exclusivity is not uncommon in the population from which the data were drawn. 
\end{abstract}

\section{Introduction}
This is a report of some work I have done under the auspices of the  the Rabadan Lab in the department of Systems Biology at Columbia University \\(https://rabadan.c2b2.columbia.edu/).
Drs. Rabadan and Camara kindly provided me with some genomic data on glioblastoma (GBM). (See section \ref{S:data}.) I dichotomized those data and applied the ``concurrence topology (CT)'' (Ellis and Klein \cite{spEaK14.ConcurTopolfMRI}) method to them. 

Concurrence topology (CT) is a method of ``topological data analysis'' that uses persistent homology to describe aspects of the statistical dependence among binary variables. I detected some one-dimensional homology with apparently long lifespan. The first question I investigated was, is that lifespan statistically significantly long? 

``Long'' compared to what? Statistical significance is always based on a ``null'' hypothesis. I took the null hypothesis to be that the observed persistent homology among mutations can be explained simply by the first and second order dependence among the mutations. Second order dependence is that which can be described fully by looking at genes just two at a time. A $p$-value can be computed by simulating data whose distribution is completely determined by the first and second order statistics of the GBM data. Specifically, I endeavored to simulate binary data sets of the size of the GBM data in such a way that all such data sets whose first and second order statistics approximate those of the GBM data are equally likely. What I mean by ``approximate'' is specified in section \ref{S:simulation}. Simulating such data is itself rather challenging (section \ref{S:simulation}) and I am not sure that my efforts were completely successful.

\section{Data and CT analysis} \label{S:data}
The GBM data set consists of data on 290 tumors. Dr. Camara recommended 75 genes of which I was able to locate 69 in the data set.  Each entry in the $290 \times 69$ matrix is a numerical score ranging from 0 to 4, inclusive. I dichotomized the data by converting every positive value to 1. So ``1'' indicates the presence of a mutation. The following table lists for every gene the number of tumors in which it was mutated.

    \begin{Small}
        \begin{equation*}
            \begin{matrix}
PTEN & TP53 & EGFR & PIK3R1 & NF1 \\
                90 & 84 & 77 & 33 & 32 \\
                PIK3CA & RB1 & MUC17 & HMCN12 & ATRX\\
                32 & 25 & 23 & 19 & 17 \\
                IDH1 & KEL & COL6A3 & STAG25 & GABRA6 \\
                15 & 15 & 14 & 12 & 11 \\
                LZTR1 & PIK3C2G & SEMG1 & F5 & RPL5 \\
                10 & 9 & 9 & 9 & 8 \\
                TPTE22 & NUP210L & IL4R & BCOR0 & BRAF\\
                8 & 8 & 8 & 7 & 6 \\
                TP63 & TRPA1 & TLR6 & QKI & PTPN11 \\
                6 & 5 & 5 & 5 & 5 \\
                PLCG1 & SETD22 & FAM126B & ZDHHC46 & TCF12 \\
                5 & 5 & 4 & 4 & 4 \\
                DDX5 & SLC6A3 & CLCN7 & RNF16818 & GLT8D2 \\
                4 & 4 & 4 & 4 & 4 \\
                TGFA & EEF1A1 & AOX1& ACAN& NIPBL \\
                4 & 4 & 4 & 4 & 3 \\
                ZNF292 & KRT13 & RBBP6 & EPHA3 & CLIP1 \\
                3 & 3 & 3 & 3 & 3 \\
                KRT15 & CREBZF & MAX & ST3GAL62 & ARID1A \\
                2 & 2 & 2 & 2 & 2 \\
                KRAS & C15orf48 & TYRP1 & ARID228 & PPM1J2\\
                2 & 2 & 2 & 2 & 2 \\
                ZBTB20 & NRAS & IL1RL1 & C10orf76 & EIF1AX \\
                1 & 1 & 1 & 1 & 1 \\
                CIC & SMARCA4 & ABCD1& EDAR
            \end{matrix}
        \end{equation*}
    \end{Small}

\begin{figure}
		      \epsfig{file = 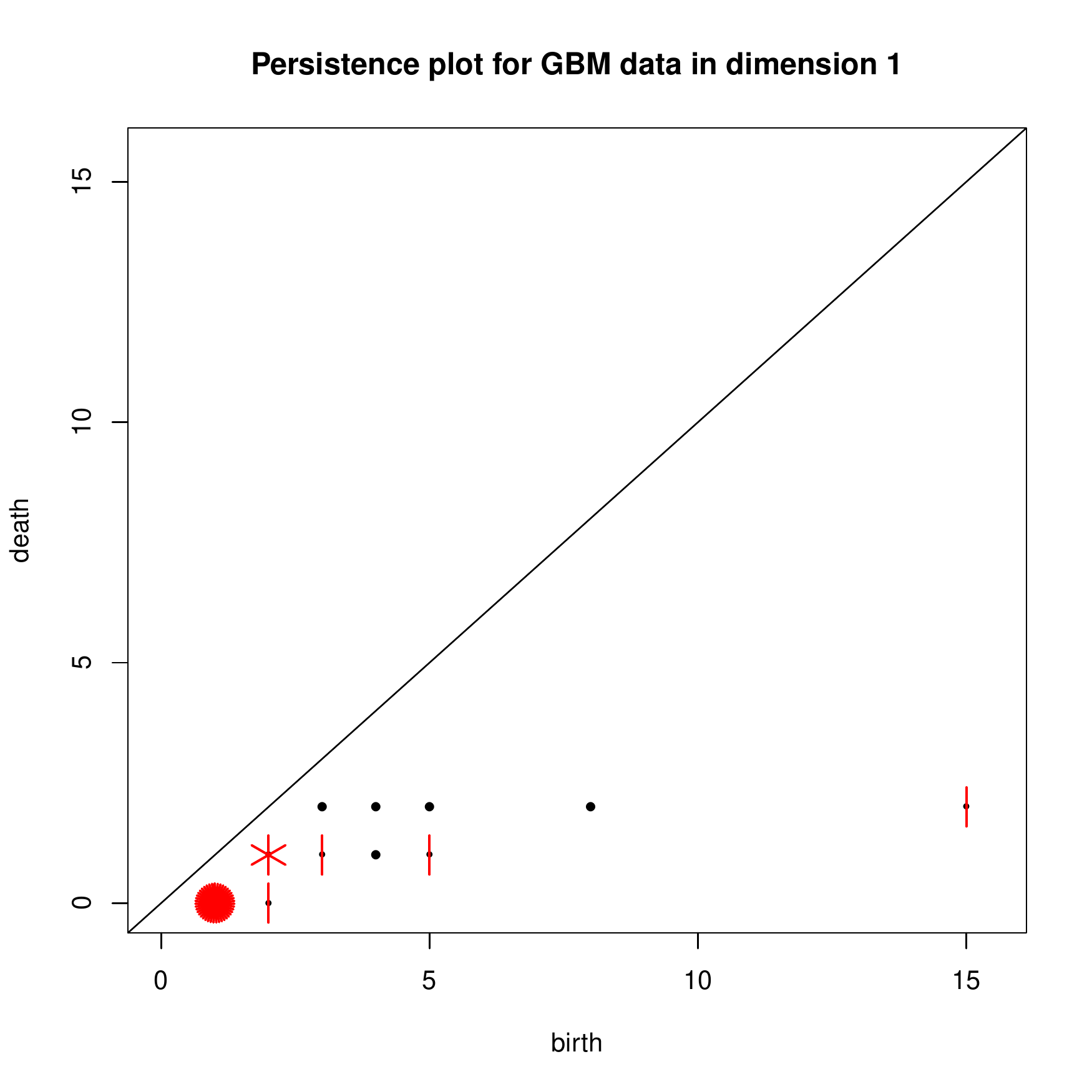, height = 5in, , }  
		      \caption{CT persistence plot in dimension 1 for GBM data. (Since CT represents data as a descending filtration, persistence is indicated by points \emph{below} the main diagonal.) In order to represent multiplicity, I employ a ``sunflower plot''. The large disk represents 43 classes born at frequency level 1. A simple dot represents one persistent class. Each ray coming out of a point represents one class with birth and death given by the coordinates at that point. For example, six classes were born at frequency level 2 and died at frequency level 1. Two classes were born at frequency level 15 and died at frequency level 3.}  
		      \label{F:GBM_persistence_plot}
	\end{figure}

I ran my CT code on this binary data set for dimension 1 and with $\mathbb{Z}/2 \mathbb{Z}$ coefficients. Figure \ref{F:GBM_persistence_plot} shows the persistence diagram.

The two persistence classes corresponding to the dot in the figure lying furthest below the diagonal line are the ones with the longest lifespan. They are born in ``frequency level'' 15 and have lifespan 13. Since at least three subjects (tumors) are needed to form a 1-cycle, each of these persistent classes involves  at least 3*15 tumors, or about 15.5\% of the sample. 

In CT it is often possible to ``localize'' classes in the sense of looking for representative ``short cycles''. A ``short cycle'' in dimension 1 means a cycle consisting of three 1-simplices, line segments. A short cycle is uniquely determined by its vertices. In this context, vertices correspond to genes. We find that each of the two classes with lifespan 13 has a short representative in frequency levels 15 and lower. They are:
     \begin{quote}
         EGFR, TP53, PTEN and \\
         PTEN, RB1, TP53.
     \end{quote}
Each of these triplets exhibit ``mutual exclusivity'' (Ciriello et al \cite{gCeCcSnS2012.MutualExclusOncogenes}, Szczurek1 and Niko Beerenwinkel \cite{eSnB.MutualExclusCancerGenes}, and Melamed et al \cite{rdMjWaIrR.GeneAlterationsGlioblastoma}). 

Note that the mutations involved in these short cycles are the three most common in the sample, EGFR, TP53, PTEN, and the seventh most common, RB1. Later we will find more evidence that there is something special about RB1.

These triples of genes reflect more than just mutual exclusivity. The corresponding 1-cycles represent homology and homology is a \emph{global} property involving all 69 genes. So we have found an apparent global pattern of mutation in GBM. However, I do not have a biological interpretation of this kind of structure. 

\section{Simulation} \label{S:simulation}
My goal was to generate random data sets that share the same first and second order statistics as the real data, at least approximately. For the cancer data this seems a little tricky. Here we discuss the algorithm I used.

Let $D$ be the $N \times d$ data matrix. So $N = 290$ is the number of samples and $d = 69$ is the number of genes. We have $N > d$. $D$ is a binary (0 -- 1) matrix and its $(i,j)$ entry is 1 if and only if the $j^{th}$ gene for the $i^{th}$ tumor is mutated. The first and second order statistics are captured by the $d \times d$ matrix $C := D^{T} D$, where ``${}^{T}$'' indicates matrix transposition. For $i \neq j$, the $(i,j)$ entry of $C$ is the number of times mutations of both genes $i$ and $j$ are present in the same sample, a second order statistic. The $i^{th}$ diagonal element of $C$ is the number of samples in which the $i^{th}$ gene is mutated, a first order statistic. 

For $n = 1, 2, \ldots$, let $1_{n}$ be the column vector all of whose entries are 1. Then $s := 1_{N}^{T} D$ is the $d$-dimensional row vector column sums of $D$. Thus, $s$ is the same as the diagonal of $C$. ``Center'' $D$ by subtracting out the column means: 
$D_{0} := D - M$ where $M^{N \times d} := N^{-1} 1_{N} \, s$. (I use superscripts to indicate matrix dimensions.) Thus, the column sums of $D_{0}$ are all 0: $1_{N}^{T} D_{0} = 0$.

Let $D_{0} = U \Lambda V^{T}$ be the singular value decomposition of $D_{0}$ (Wikipedia). Thus, $U$ is an $N \times d$ matrix with orthonormal columns, $\Lambda$ is $d \times d$ diagonal, and $V$ is orthogonal. For our data sets no entry of $\Lambda$ is 0 and all the entries are distinct. Since $0 = 1_{N}^{T} D_{0} = 1_{N}^{T} U \Lambda V^{T}$ it follows that 
    \begin{equation}  \label{E:1.N.U.=.0}
        1_{N}^{T} U = 0.
    \end{equation} 

Moreover, $D_{0}^{T} D_{0} = V \Lambda^{2} V^{T}$. Hence, the diagonal of $\Lambda^{2}$ is the vector of eigenvalues of $D_{0}^{T} D_{0}$ and the columns of $V$ are the corresponding unit eigenvectors. Therefore, since the eigenvalues are distinct, the columns of $V$ are unique up to sign. 
Observe that $N^{-1} D_{0}^{T} D_{0}$ is just the (variance-)covariance matrix of $D$. Hence, columns of $V$ are the unit eigenvectors of the covariance matrix and the diagonal elements of $N^{-1} \Lambda^{2}$ are the eigenvalues.

We have
    \begin{equation} \label{E:D=U.Lamb.VT.+.M}
        D = D_{0} + M = U \Lambda V^{T} + M
    \end{equation}
where, you recall, $D$ is the original data matrix. Let $W^{N \times 1}$ be any matrix with orthogonal columns s.t.\ (such that) $1_{N}^{T} W = 0$. For example, by \eqref{E:1.N.U.=.0}, we can take $W = U$. Let $Y := W  \Lambda V^{T} + M$. Since $1_{N}^{T} W = 0$ we have 
$W^{T} M = N^{-1} W^{T} 1_{N} s = 0$. Similarly, $M^{T} M = N^{-2} s^{T} 1_{N}^{T} 1_{N} s = N^{-1} s^{T} s$, 
since $1_{N}^{T} 1_{N}  = N$.  Thus, since the columns of $W$ are orthonormal, 
    \begin{equation*}
        Y^{T} Y = ( V \Lambda W^{T} + M^{T} ) (W  \Lambda V^{T} + M) = V \Lambda^{2} V^{T} + N^{-1} s^{T} s.
    \end{equation*}
So $Y^{T} Y$ \emph{does not depend on} $W$. In particular, $Y^{T} Y = D^{T} D = C$. 

We can use this fact to sample uniformly from the set of all $N \times d$ matrices, $Y$, s.t. $Y^{T} Y = C$, i.e. to sample uniformly from the set of all $N \times d$ matrices having the same first and second order statistics that $D$ has. One merely has to sample uniformly from the space of all matrices $W^{N \times 1}$ with orthogonal columns s.t.\ (such that) $1_{N}^{T} W = 0$. 

Such sampling can be done easily as follows. Let $w^{N \times 1}$ be a random Gaussian column vector with statistically independent \emph{population} mean 0 components. Let $\bar{w} = N^{-1} 1_{N}^{T} w$ be the \emph{sample} mean of the components of $w$. ($\bar{w}$ is just a random number.) Center $w$, i.e., replace it by $w_{1} := w -  \bar{w} 1_{N}$. 
Thus, $w_{1}$ is a random $N$-column vector and $1_{N}^{T} w_{1} = 0$. Repeat this operation independently $d$ times producing column vectors $w_{1}, \ldots, w_{d}$. Apply the Gram-Schmidt orthogonalization process to these vectors to produce orthonormal column $N$-vectors $w_{1}', \ldots, w_{d}'$. Since $w_{1}, \ldots, w_{d}$ all have sample mean 0, so do $w_{1}', \ldots, w_{d}'$. Let $W^{N \times d}$ be the matrix whose columns are $w_{1}', \ldots, w_{d}'$. Finally, take $Y := W  \Lambda V^{T} + M$. Then we know $Y^{T} Y = C$.

I mentioned above that the columns of $V$ are unique up to sign.  One might think that one can make the distribution of $Y$ more uniform by randomly changing the signs of the columns of $V$. Let $E^{d \times d}$ be a diagonal matrix with diagonal entries, $\epsilon_{i} = \pm 1$, ($i=1, \ldots, d$). 
Since diagonal matrices commute, we have $W  \Lambda (V E)^{T} = W  \Lambda E V^{T} = (W E)  \Lambda V^{T}$. Thus, multiplying the columns of $V$ 
by $\epsilon_{i}$, ($i=1, \ldots, d$) amounts to replacing $w_{1}', \ldots, w_{d}'$ by $\epsilon_{1} w_{1}', \ldots, \epsilon_{d} w_{d}'$. Now, examination of the Gram-Schmidt process shows that $\epsilon_{1} w_{1}', \ldots, \epsilon_{d} w_{d}'$ is exactly what one gets when one applies Gram-Schmidt to the original $\epsilon_{1} w_{1}, \ldots, \epsilon_{d} w_{d}$, i.e.\ changing the signs of the independent Gaussian random vectors we started with. But theses random vectors are independent Gaussian with mean 0, therefore changing their signs does not change their distribution. This shows that changing the signs of the columns of $V$ does not change the distribution of $Y$ and, thus, is unnecessary.

$Y$ constructed as above is uniformly distributed over the space of matrices $X^{N \times d}$ with $X^{T} X = C$. There is only one problem. We want \emph{binary} matrices and the probability that $Y$ generated as above is binary is 0. The obvious remedy is to threshold: Write $Y = ( y_{ij} )$. For some number $t$ replace each entry $y_{ij}$ by 0 if $y_{ij} < t$ and by 1 otherwise. Call the resulting binary matrix $B_{t}^{N \times d}$. But what threshold $t$ should we use?

Alas, we have to accept the fact that no matter what $t$ we use we will \emph{not} have $B_{t}^{T} B_{t} = C$. So we have to settle for an approximation $B_{t}^{T} B_{t} \approx C$. We pick $t$ to get a ``best'' approximation.  

In order to define what ``best'' means we need a definition of distance between $B_{t}^{T} B_{t}$ and $C$. One possibility is to use the (squared) Euclidean or Frobenius matrix distance: $trace \, (B_{t}^{T} B_{t} - C)^{T} (B_{t}^{T} B_{t} - C)$. This is just the sum of the squared entries of $B_{t}^{T} B_{t} - C$. But remember that $D$, and hence, $C$ are random and a more stable distance would use weights approximately equal to the variances of the entries of $C$. 

There are numerous ways might estimate these variances. I employed a simple one. Remember that the entries of $C$ are counts, non-negative integer-values. Perhaps the simplest distribution of a non-negative integer-valued random variable is the Poisson. So a crude estimate of the variance of an entry is just the entry itself. However, there are hundreds of unique values in $C$ so the values themselves will be very noisy estimates. 

Now, in statistics it is well known that when simultaneously estimating a large number of quantities one improves estimates by shrinking toward a constant. I informally employed that technique. Let $c$, a number, be the sample mean of all the entries in $C$ and let $\bar{C}$ be the $d \times d$ matrix all of whose entries are $c$. Then for the purpose of weighting we replace $C$ by $\hat{C} := (1/2) C + (1/2) \bar{C}$. 

Now define the ``distance'', $\delta(B_{t}^{T} B_{t}, C)$, between $B_{t}^{T} B_{t}$ and $C$ as follows. Form the matrix $\Delta_{2}^{d \times d} = (\delta_{ij})$ whose $ij^{th}$ entry is the squared difference between the $ij^{th}$ entry of $B_{t}^{T} B_{t}$ and the $ij^{th}$ entry of $C$. Now divide $\delta_{ij}$ by the $ij^{th}$ entry of $\hat{C}$. Add up all those quotients. That's the ``distance'', $\delta(B_{t}^{T} B_{t}, C)$. However, $B_{t}^{T} B_{t}$ is symmetric, so in this procedure the unique off-diagonal elements are counted twice. To make up for this, we modify $\hat{C}$ by dividing its diagonal by 2. This doubles the contribution of the diagonals of $B_{t}^{T} B_{t}$ and $C$.

The binary matrix we want will not be all 0 or 1. Therefore, the only thresholds we need try are the distinct numeric values in $Y$. We try all those values as thresholds and pick the one that minimizes $\delta(B_{t}^{T} B_{t}, C)$. 

How do we know that the minimum distance we achieve is small enough? Again, the data matrix $D$ itself is random. Even if we gathered another data set $D_{2}$ using the same method used to gather $D$ we would not have $D_{2}^{T} D_{2} = C$. Therefore, it is unreasonable to insist that $\delta(B_{t}^{T} B_{t}, C)$ be tiny.  

But how small a value for $\delta(B_{t}^{T} B_{t}, C)$ is acceptable? Looking at the distribution of $\delta(D_{2}^{T} D_{2}, C)$, where $D_{2}$ is drawn at random from the same population that $D$ came from gives us a yardstick to use for judging sizes 
of $\delta(B_{t}^{T} B_{t}, C)$. Now, we cannot draw new samples $D_{2}$, but we can approximate that process by drawing samples, with replacement, from the rows of $D$. This is the non-parametric bootstrap (Efron and Tibshirani \cite{bErjT93.bootstrap}). One draws many samples, with replacement, from the rows of $D$. (These are called ``resamples''. I drew 2,000.) Each time one obtains a matrix $D_{2}$. One then records the value of $\delta(D_{2}^{T} D_{2}, C)$ for each resample. The distribution of all these numbers approximates the distribution one would get by taking many samples $D_{2}$ from the population. I chose the median $m_{2}$ of these distances as the cutoff for distinguishing matrices that are close to $D$ from those that are not. 

Unfortunately, even the closest $B_{t}$, call it $B_{t_{opt}}$, generated by thresholding $Y$ practically always fails this test. 
So more work needs to be done to $B_{t_{opt}}$ to make it acceptable. To do this I used an informal ``Markov Chain Monte Carlo (MCMC)'' algorithm (Wikipedia).  Intialize $Z_{1} := B_{t_{opt}}$ and $b_{1} := \delta(Z_{1}^{T} Z_{1}, C)$. At each step pick a random entry in $Z_{1}$ and flip it so 0 gets replaced by 1 or vice versa. (Call that a ``flip attempt''.) Call the resulting matrix $Z$. Then compute $b := \delta(Z^{T} Z, C)$. If $b < b_{1}$ (a ``successful flip'') then set $Z_{1} := Z$ and $b_{1} := b$. Otherwise, $Z_{1}$ and $b_{1}$ are not changed. 

That is the iteration. What is the stopping rule? I stopped the iteration as soon as $b_{1} < m_{2}$. One might be concerned that when the algorithm halts the distance $b_{1} := \delta(Z_{1}^{T} Z_{1}, C)$ would be only slightly smaller than $m_{2}$. Values much smaller than $m_{2}$ would never be achieved. However, it is well known that the volume of a high dimensional ball is almost entirely found near the boundary. So if one did sample matrices from the ball centered at $D$ and having $\delta$-radius $m_{2}$, one would rarely get a matrix whose $\delta$-distance from $D$ is much smaller than $m_{2}$. 

So that is not a legitimate objection to the ``MCMC'' algorithm. A more justified concern is the following. Above we argued that $Y$ constructed as above is uniformly distributed over the space of matrices $X^{N \times d}$ with $X^{T} X = C$. Even the matrix 
$B_{t_{opt}}$, though not close to $D$, is unbiased in terms of the \emph{direction} $B_{t_{opt}} - D$. I.e., I conjecture that it would not be hard to prove that the expected value of $B_{t_{opt}} - D$ is 0. However, I fear that the informal MCMC step in the construction might introduce some bias.  

Still, why not skip the SVD step and just perform the ``MCMC'' step? As an experiment I generated a 0 matrix of the same dimensions as the data and randomly replaced 712 of the entries by 1's, where 712 is the number of 1's in the original data matrix.. It took more than 21,000 flip attempts with over 1,000 successful flips to bring this matrix close to the data matrix. A second attempt produced similar results. As a benchmark, note that there are 20010 positions in data matrix. But it is not just to save flips that the svd-based starting matrix is helpful. That approach also serves, I believe, to help generate binary matrices (nearly?) uniformly distributed among matrices with first and second order statistics similar to the real population.

\section{Simulation results and bootstrap} \label{S:results}
I generated 500 synthetic data sets using the algorithm described in section \ref{S:simulation}. In contrast to the experiments described in the last paragraph of the last section, in these 500 MCMC calculations, the range in the number of flip attempts needed to bring the second order statistics acceptably close to those of the data was 625 to 2076 with a median of 1061. The number of successful flips ranged from 193 to 329 with a median of 255.

For each synthetic data set I found the persistent 1-D homology classes with the longest lifespan. (There were practically always just one such class.) Here are summary statistics for those lifespans:
    \begin{equation*}
        \begin{matrix}
     Min. & 1st Qu. & Median  & Mean & 3rd Qu. & Max.  \\
     5.00  & 12.00  & 14.00  & 14.36  & 17.00  & 25.00
        \end{matrix}
    \end{equation*}
We observe that the maximum lifespan in the real data, viz., 13, is not remarkable in the simulated data. In fact, 56.4\% of the time the maximum lifespan obtained in simulation was larger than in the data.

For those classes with maximum lifespan I also recorded their frequency level of birth. Here are summaries.
    \begin{equation*}
        \begin{matrix}
    Min. & 1st Qu. &  Median  & Mean & 3rd Qu. & Max.  \\  
     6.00  & 19.00  & 22.00  & 20.63  & 24.00  & 32.00
        \end{matrix}
    \end{equation*}
We observe that the births in the real data, viz., 15, is actually rather small by comparison. In fact 84.4\% of the time the birth of the class with maximum lifespan obtained in simulation was larger than in the data.

The vertices that appear in one or both of the short cycles I found in the longest lived classes in the GBM data (section \ref{S:data}) are EGFR, PTEN, RB1, and TP53. For each of them I recorded if it appeared in a short cycle in the simulated data. (The persistent class with the longest lifespan was represented by one or more short cycles in all but ten of the simulations.) Here are the proportions of the simulations in which that happened. 
    \begin{equation*}
        \begin{matrix}
EGFR & PTEN & RB1 & TP53 \\
     0.914  &   0.968  &   0.080   &  0.970
        \end{matrix}
    \end{equation*}
We see, then, that the homology class represented by the short cycle with vertices PTEN, RB1, TP53 appears rather uncommonly in the simulated data (because RB1 appears uncommonly).

But perhaps the cycle with vertices PTEN, RB1, TP53 also appears uncommonly in the population from which the GBM data is derived. To answer this question, in the spirit of Chazal et al \cite{fCbtFfLaRaSlW14.BootPersDiags}, I applied the bootstrap method mentioned in section \ref{S:simulation} with 500 resamples. Thus, I resampled the tumors (rows) of the GBM matrix and computed the same summaries I just described for the SVD simulations. I found that in over 25\% of the resamples the longest lived classes had short cycles including RB1 as a vertex. This is  (circumstantial) evidence that in the population of tumors from which the data are drawn the homology class represented by the short cycle with vertices PTEN, RB1, TP53 is not uncommon.

\section{Discussion} \label{S:discussion}
Applying the Concurrence Topology method to the GBM data we found two cycles with apparently long lifespan. But ``long'' compared to what? Compared to results one would get just ``by luck''. But what kind of luck? Complete independence among mutations is biologically unrealistic. It is already known that mutations are not independent. (See, for example, the afore-cited articles on mutual exclusivity.) Instead, I focussed on the luck one would observe if the distribution of mutations reflected, approximately, the first and second order statistics in the data. 

I used an algorithm that is intended to generate samples from such a distribution. It is not hard to generate \emph{floating point}-valued matrices whose first and second order statistics exactly match those of the data. The difficulty is in satisfying the requirement that the algorithm produce \emph{binary} matrices having the desired distribution. Specifically, one would like all binary matrices having approximately the right statistics (where ``approximately'' is defined in section \ref{S:simulation}) to be equally likely. The method I used takes the floating point algorithm as starting point then discretizes and randomly flips entries to achieve an approximate match. Perhaps better algorithms already exist somewhere, otherwise more work in this area is needed.

We found that neither the frequency level of birth nor lifespan of the classes in the real data is remarkable in that second order context. However, one of the classes with the longest lifespan in the data appears rather infrequently in the second order distribution but not infrequently in bootstrapped resamples. This suggest the involvement of third and higher order dependence among the mutations.

I find it surprising that first and second order dependence can give rise to persistent homology in dimension 1 with long lifespans. I think this is partly due to the fact that some of the mutations, \emph{viz.}, EGFR, TP53, PTEN, are so common. 

In section \ref{S:data} I pointed out that persistent homology reflects ``mutual exclusivity''.  But mutual exclusivity is a ``local phenomenon'': mutual exclusivity among a group of mutations, e.g.\ EGFR, TP53, PTEN, is a property of the joint distribution of just those mutations. But persistent homology is a \emph{global} property. In general, it reflects the joint distribution of all, in this case 69, genes. In the case of the short cycle EGFR, TP53, PTEN the fact that it represents persistent homology with a long lifespan may just be local because those mutations are far more numerous than the others. However, the short cycle PTEN, RB1, TP53 seems to reflect something more global because RB1 is not such a common mutation. The simulations seem to confirm this.

\newcommand{\etalchar}[1]{$^{#1}$}
\providecommand{\bysame}{\leavevmode\hbox to3em{\hrulefill}\thinspace}
\providecommand{\MR}{\relax\ifhmode\unskip\space\fi MR }
\providecommand{\MRhref}[2]{%
  \href{http://www.ams.org/mathscinet-getitem?mr=#1}{#2}
}
\providecommand{\href}[2]{#2}

\end{document}